\documentclass[runningheads, orivec]{llncs}
\usepackage[T1]{fontenc}
\usepackage{tikz}
\usetikzlibrary{shapes.geometric} 
\usepackage{array}
\newcolumntype{R}[1]{>{\raggedleft\let\newline\\\arraybackslash\hspace{0pt}}p{#1}}
\usepackage{amsmath}
\usepackage{cases}
\usepackage[implicit=true,pdftex,colorlinks=true,urlcolor=blue,linkcolor={red!50!black},citecolor={blue!50!black}]{hyperref}
\usepackage{etoolbox}
\usepackage[ruled,vlined]{algorithm2e}
\SetKwComment{mcc}{//}{}
\appto\UrlBreaks{\do\-}
\usepackage{microtype}
\usepackage{color}

\usepackage{bbold,bm}
\usepackage{accents}
\newcommand\norm[1]{\lambda\left( #1 \right)}
\newcommand\dist[1]{\left| #1 \right|_2}
\newcommand{\olsi}[1]{\,\overline{\!{#1}}} % overline short italic
\newcommand{\bu}{\bm{u}}
\newcommand{\bs}{\bm{s}}
\newcommand{\bx}{\bm{x}}
\newcommand{\F}{\mathcal{F}}
\newcommand{\N}{\mathcal{N}}
\newcommand{\U}{\mathcal{U}}
\renewcommand{\S}{\mathcal{S}}
\newcommand{\T}{\mathcal{T}}
\newcommand{\Zsim}{\sim_{\scriptscriptstyle Z}}
\newcommand{\ZS}{\S/\mathclose{\Zsim}\mathopen{}}
\newcommand{\region}{\mathcal{R}}
\newcommand{\Tregion}{\olsi{\mathcal{R}}}
\newcommand{\NS}{\N_{\S}}
\newcommand{\NSZ}{\N_{\ZS}}
\newcommand{\ONS}{\olsi{\N}_{\S}}
\newcommand{\RI}{\textsf{R}^{\textsf{I}}_{\S}}
\newcommand{\TRI}{\accentset{\sim}{\textsf{R}}^{\textsf{I}}_{k,\delta}}

\begin{document}
\title{Disclosure risk in a geo-spatial setting}
\author{Peter-Paul de Wolf%\inst{1}\orcidID{0000-0003-3953-2410} \and
%Second Author\inst{2,3}\orcidID{1111-2222-3333-4444} \and
%Third Author\inst{3}\orcidID{2222--3333-4444-5555}
}
\authorrunning{P.P. de Wolf}
\institute{Statistics Netherlands, The Hague, The Netherlands\footnote[1]{The views expressed in this paper are those of the author and do not necessarily reflect the policy of Statistics Netherlands.\\
The author likes to thank Edwin de Jonge and Rosanne Turner for stimulating discussions on this topic and for reviewing an earlier version of this paper.}\\
\email{pp [dot] dewolf [at] cbs [dot] nl}
%\and
%Springer Heidelberg, Tiergartenstr. 17, 69121 Heidelberg, Germany
}
\maketitle              % typeset the header of the contribution
\begin{abstract}
Using thematic maps to publish statistical information has become a popular visualization. As is the case with all statistical publications, thematic maps also have to deal with the balance between disclosure risk and utility. However, most risk and utility measures do not take into account the spatial character of a map. Some of the proposed spatial risk measures suffer from the Modifiable Areal Unit Problem (MAUP): slightly changing regional classifications may influence the risk. Indeed, even a small translation of for example a grid may influence that risk. We propose a new risk measure that does not suffer from MAUP. Moreover, our risk is directly related to the local density of the (target) population and takes into account that often multiple units may be connected to a single location. We show the behavior of our risk measure using an example dataset of fake but realistic locations of enterprises. Our risk measure can be adapted to take into account the effect on the (perceived) risk of zooming in or out and the effect of the used resolution.
\end{abstract}
\section{Introduction}
Users of statistical information are often interested in determining whether the value of a statistic depends on spatial patterns. Policymakers, for example, may want to identify which areas of their cities require special attention in their particular policy domain.
Additionally, visualizing spatial patterns can help communicate results to a wider audience. Humans are visually oriented and often prefer nice visualizations over dry tabulations of mere numbers.

As with any statistical output, publishing information using visualizations has to deal with the tension between confidentiality and utility. 
Spatial data can be very intrusive and is often considered to belong to the most relevant information for disclosure. 
Indeed, geographic information can reveal the location of sensitive infrastructure, potentially compromising national security. Additionally, spatial data can be highly identifying, enabling the attribution of characteristics to specific units at known locations.

Dealing with confidentiality in publishing statistical information is the subject of the research area of Statistical Disclosure Control (SDC). See for example \cite{Hundepool2012book} and \cite{Hundepooletal2026} for an overview of general SDC literature. For information on SDC targeted at geo-referenced data, see \cite{Mohler2025}.

In this paper we will focus on disclosure risk when publishing statistical data on a cartographic map. With a cartographic map we mean a visual map that relates two-dimensional physical locations to units in a (target) population. More specifically, we will focus on \emph{thematic} maps: cartographic maps that display some (non-spatial) attribute of the units associated with their locations. For example, maps that display energy used by enterprises or that display the social economic status of individuals.

Thematic maps can come in several flavors, such as point maps, choropleth maps, dot maps, density maps, gridded maps, heat maps, referring to the way the information is displayed. Traditionally, thematic maps are constructed using data to which SDC has already been applied, either at microdata level or at aggregated level, depending on the flavor of the used mapping. However, often SDC methods applied to microdata or tabular (aggregated) data do not take into account the actual spatial character of the data. Neither in calculating the risk nor in applying protection methods.

In previous work (see e.g., \cite{JongeWolf2016}, \cite{WolfJonge2017}, \cite{WolfJonge2018} and \cite{Hut2020}) we discussed some aspects of SDC applied to thematic maps. We proposed some risk measures based on measures commonly used in tabular data protection. These measures however did have some shortcomings. In the current paper we propose a new risk measure that overcomes some of these shortcomings.

In particular we propose a risk measure that does not suffer from the so called MAUP (Modifiable Areal Unit Problem) where a mere translation of a region influences the value of the risk measure. Moreover, we think that our newly proposed measure is more suitable for density maps and heat maps. And last but not least, we think that our measure can easily be adjusted to  capture effects on the (perceived) risk of different zoom factors and resolutions.

In section~\ref{sec:risk_general} we will first introduce the notation that we will use throughout the paper. We will also mention some basic concepts related to spatial data and we will illustrate the MAUP idea.
The next section~\ref{sec:altmeasure} introduces the alternative risk measure we propose. Section~\ref{sec:example} gives an application of the newly introduced measure for a dataset on enterprise locations. Finally, in section~\ref{sec:discussion} we conclude with a small discussion on the measure and the outcome of the application of the measure to the example.

\section{Disclosure risk and geography}\label{sec:risk_general}
Disclosure risk is often defined as a combination of the probability of occurrence of a disclosure and the impact when such disclosure happens. 
However, the term disclosure risk is often used only as the probability of occurrence. 
In the current paper we will also concentrate on that aspect and thus, implicitly, assume that any disclosure has the same impact. 
This may be accomplished by calculating the risk only in case sensitive information (high-impact information) is involved.

\subsection{Notation and basic concepts}
Obviously, location is a basic concept in spatial analysis. Thematic maps display (statistical) information about units associated with locations or regions. In this situation it is convenient to represent these units by their locations and we will thus denote the population by $\U = \{ \bu_1,\ldots,\bu_N \}$ with $\bu_j = (x_j,y_j)$ the location of unit $j$.
However, multiple units may be associated with the same location, e.g., when two people live at the same address, or when two people live `above' each other in  an apartment building. 
Hence, $\U$ could be a multiset: a set in which the same element may occur multiple times. For the purpose of the current paper, we can view such a multiset as a set $\S = \{ \bs_1,\ldots,\bs_M \}$ with $\bs_i = (x_i, y_i, m_i)$, the $xy$-coordinates in $\mathbb{R}^2$ of unit $i$ together with the multiplicity $m_i \in \mathbb{N}$. The total population size is thus $N = \sum_{i=1}^M m_i$.
Note that this situation also happens in case of gridded maps where each grid cell is considered as a single location (e.g., its center-point) with all units within that cell assigned to that single location. 
Moreover, from a visualization point of view, zooming out may result in seemingly overlapping locations. In that situation the \emph{displayed} locations can also have a multiplicity larger than one. This is a visualization effect that lowers the perceived risk of identification. For more information on perceived disclosure risk in relation to maps, see e.g., \cite{Kim2020}.

Let $\region$ denote a region, i.e., a non-empty connected open subset of $\mathbb{R}^2$. 
Let $\NS(\region)$ count the number of units $\bu_j$ of population $\U$ (related to the unique locations $\bs_i$ in $\S$) that are located in region $\region$:
\begin{equation}
    \NS(\region) = \sum_{i=1}^M m_i\mathbb{1}\big((x_i,y_i) \in \region\big)
\end{equation}
where $(x_i,y_i)$ are the location coordinates of $\bs_i$, $m_i$ is the multiplicity of $\bs_i$ and $\mathbb{1}(A)$ equals one if $A$ is true and zero if $A$ is false. Note that $\NS(\mathbb{R}^2) = N$, the total number of units in $\U$.

The areal number density is the number of population units per unit area:
\begin{equation}
    {\ONS}(\region) = \frac{\NS(\region)}{\norm{\region}}
\end{equation}
where $\norm{\region}$ is the two-dimensional Lebesgue measure of region $\region$ (the `spatial size' or `area' of region $\region$).

\subsection{Areal identity risk and MAUP}
In \cite{WolfJonge2017}, for regions with $\NS(\region) \neq 0$ we defined the areal identity disclosure risk as
\begin{equation}\label{eq:RI}
    \RI(\region) = \begin{cases}
                        1\big/\NS(\region) & \textrm{if }\NS(\region) \neq 0 \\
                        0 & \textrm{if }\NS(\region) = 0
                   \end{cases}
\end{equation}
Intuitively this measure means that, given the region $\region$, the risk of identification is smaller the more units are inside that region, with the exception that for a region without any units the areal identity risk is set to zero.
Note that $\RI(\cdot)$ has the property that for non-empty regions $\region_1$ and $\region_2$ such that $\region_2 \subseteq \region_1$ we get $\RI(\region_1) \leq \RI(\region_2)$, i.e., enlarging the size of the region will not increase the areal identity disclosure risk. In particular, as shown in Figure~\ref{fig:example1}, when the number of units is fixed, enlarging the area will not change the risk: $\NS(\region_1) = \NS(\Tregion_1) = 1/3$.
One might argue that the risk of the units at $\bs_1$, $\bs_2$ and $\bs_3$ in Figure~\ref{fig:example1} is higher when considering $\region_1$ compared to $\Tregion_1$: these units look more isolated (as a group) and hence might be easier to identify.

This measure is subject to the so called Modifiable Areal Unit Problem (MAUP). This problem was first recognized in \cite{Gehlke_Biehl1934} and later described in more detail in \cite{openshaw1983}. This MAUP essentially states that results may differ during spatial analysis of aggregated data depending on the used regions. MAUP can take two forms: a scaling effect and a shape effect. One instance of a shape effect for this risk measure is visible in Figure~\ref{fig:example2}: slightly translating a region affects the value of the risk measure. Indeed, $\NS(\region_2) = 1/3$ but $\NS(\Tregion_2) = 1/2$. Figure~\ref{fig:shapeexample} shows the MAUP effect by changing a square region into a hexagonal shape of the same size: then we get $\NS(\region_3) = 1/3$ but $\NS(\Tregion_3) = 1/4$. In some sense, one could thus regard $\RI(\region)$ to measure the risk of the \emph{area} instead of the risk of a \emph{unit}. Indeed, that is the reason for denoting the measure as a function of $\region$.

\begin{figure}[htb]
\begin{minipage}[b]{0.5\textwidth}
\centering
\begin{tikzpicture}[scale=0.75]
    \draw (0,4) node[below left] {$\region_1$} (0,0) rectangle (4,4);
    \draw[dashed] (2,2) node[above right] {$\Tregion_1$} (2,0) rectangle (4,2);
    \fill (2.2,0.2) circle (2pt) node[right] {$\bs_1$}
          (2.9,1.7) circle (2pt) node[below] {$\bs_2$}
          (3.2,1.8) circle (2pt) node[right] {$\bs_3$}
          (-0.2,2.0) circle (2pt) node[below left] {$\bs_4$};
\end{tikzpicture}
\caption{Resized area}\label{fig:example1}
\end{minipage}%
\begin{minipage}[b]{0.5\textwidth}
\centering
\begin{tikzpicture}[scale=0.75]
    \draw (0,4) node[below left] {$\region_2$} (0,0) rectangle (4,4);
    \draw[dashed] (4.3,0.4) node[above right] {$\Tregion_2$} (0.3,0.4) rectangle (4.3,4.4);
    \fill (2.2,0.2) circle (2pt) node[right] {$\bs_1$}
          (2.9,1.7) circle (2pt) node[below] {$\bs_2$}
          (3.2,1.8) circle (2pt) node[right] {$\bs_3$}
          (-0.2,2.0) circle (2pt) node[below left] {$\bs_4$};          
\end{tikzpicture}
\caption{Translated area}\label{fig:example2}
\end{minipage}
\centering
\begin{minipage}{0.5\textwidth}
\centering
\begin{tikzpicture}[scale=0.75]
    \pgfmathsetmacro{\RP}{sqrt(32 / (3*sqrt(3)))}
    \draw (0,4) node[below left] {$\region_3$} (0,0) rectangle (4,4);
    \draw (4.3,2.0) node[right] {$\Tregion_3$};
    \node[dashed, regular polygon, regular polygon sides=6, draw, minimum size={1.5*\RP cm},rotate=0] at (2,2) {};
    \fill (2.2,0.2) circle (2pt) node[right] {$\bs_1$}
          (2.9,1.7) circle (2pt) node[below] {$\bs_2$}
          (3.2,1.8) circle (2pt) node[right] {$\bs_3$}
          (-0.2,2.0) circle (2pt) node[below left] {$\bs_4$};
\end{tikzpicture}
\caption{Differently shaped area}\label{fig:shapeexample}
\end{minipage}
\end{figure}

However, in the end the risk for individual units has to be determined. Under an attacker where the attacker only knows the region a unit belongs to, but not its exact location, choosing a unit randomly from the population within the known region $\region$ has the probability of $\RI(\region)$ to be the correct unit.

It is tempting to say that this is comparable to the situation of frequency count tables. 
For that kind of tables the sensitivity of a cell also depends on the classification of the spanning variables that define the table. 
Aggregating e.g., neighborhoods into municipalities influences the number of counts in each cell and hence the sensitivity of the cells according to a threshold rule.
However, for thematic maps this is slightly different. Thematic maps are a mixture of information on individual and aggregated level. Often it is known to which region a unit belongs; its exact location may even be known. Moreover, the thematic information can be published for a region as a whole (choropleth maps, heatmaps), or for each location individually (dot maps).

%The risk measure is defined for any arbitrary region $\region$. Ideally the risk should be calculated using the `natural' region that relates to an individual unit of the (target) population. In practice, it turns out to be difficult to define or derive such a `natural' region. Therefore it is appropriate to use a predefined set of regions to be checked for disclosure: a set of grid cells,  unions of grid cells or other `logical' regions (e.g., the region of the building where the business in question resides).

\section{Alternative measure}\label{sec:altmeasure}
In this section we will propose an alternative measure that is not susceptible to MAUP and can easily be extended to include the visualization effect of zooming. To deal with MAUP, we propose a measure that is not depending on a pre-selected area. Let $B(\bx,r)$ denote the ball in $\mathbb{R}^2$ (i.e., a solid circle) with radius $r$, centered at $\bx$. Then, for any location $\bs \in \mathbb{R}^2$ we define for some $k\in\mathbb{N}$ and $\delta \geq 0$
\begin{equation}\label{eq:Fkdelta}
    \F_{k,\delta}(\bs) = \min
    \big\{r\!: \NS(B(\bx,r))\geq k, \dist{\bx-\bs}\leq\min(r,\delta), \bx \in \mathbb{R}^2\big\}
\end{equation}
where $\dist{\bx-\bs}$ denotes the $L_2$-distance between $\bx$ and $\bs$.

A region defined by $B(\bx,r)$ with $\NS(B(\bx,r))\geq k$ we will call a \emph{$k$-anonymity region}. We thus effectively look for the smallest $k$-anonymity region that contains $\bs$ and has its center $\bx$ at a distance from $\bs$ of at most $\delta$. The larger the value of $\F_{k,\delta}(\bs)$ for fixed $k$ and $\delta$, the higher the risk: $\bs$ is more isolated when a large $k$-anonymity region is needed. 

Additionally, note that this measure is monotone non-increasing in $\delta$:
\begin{equation}
    \delta_1 \leq \delta_2 \implies \F_{k,\delta_2}(\bs) \leq \F_{k,\delta_1}(\bs)
\end{equation}
as increasing $\delta$ increases the space of admissible $\bx$.

Since the function $\NS(\T)$ counts the number of population units in region $\T$ and locations may have a multiplicity larger than 1, the minimal $k$-anonymity region may contain more than $k$ units. To correct for this, we suggest to use the following risk measure:
\begin{equation}\label{eq:Rkdelta}
    \TRI(\bs) = \frac{\F_{k,\delta}(\bs)}{\NS(B(\bx^*,r^*))}
\end{equation}
where $\bx^*$ and $r^*$ are the center and radius of the the minimal $k$-anonymity region connected to $\bs$. This combines the geographical anonymity with the local indistinguishability of population units when some thematic value is published. Note that in theory, this adjustment may destroy the monotonicity: a smaller minimal $k$-anonymity region connected to $\bs$ may contain less as well as more units than a larger $k$-anonymity region connected to $\bs$, depending on the multiplicities of the contained locations. 

When evaluating this risk for units $\bu\in\U$, the value of the parameter $\delta$ is related to prior knowledge of an attacker: the better an attacker knows the direct surroundings of a targeted unit, the smaller the $\delta$ has to be taken. Indeed, taking $\delta=0$ implies the attacker knows the exact location of the targeted unit. In that case the minimal $k$-anonymity region is best taken with the target unit in the center. Taking $\delta=\infty$ implies the attacker does not know the direct surroundings and the minimal $k$-anonymity region can be located/centered anywhere (as long as the target unit is still inside).

\subsection{Zooming effect}\label{subsec:zooming}
As discussed before, zooming can lead to the visualization effect that different units seem to overlap. Obviously, in terms of absolute coordinates they never overlap, but when zoomed out the effective coordinates are less precise and the corresponding locations effectively overlap. To formalize this zooming, define a zoom-projection as
\begin{equation}
    \pi_Z: \mathbb{R}^2 \to \mathbb{R}^2_{\textit{screen}}
\end{equation}
where $\mathbb{R}^2_{\textit{screen}}$ is the space of (screen-)pixels.

This in turn introduces the equivalence relation
\begin{equation}
    \bs_i \Zsim \bs_j \iff \pi_Z(\bs_i) = \pi_Z(\bs_j)
\end{equation}
and equivalence class $[\bs]_{\scriptscriptstyle Z}$ of all locations in $\S$ that are equivalent to $\bs$. This way we end up with a quotient space $\ZS = \pi_Z(\S)$ of equivalence classes $\bs^Z_l$ for $l=1,\ldots,L_Z$. To each equivalence class we assign the multiplicity
\begin{equation}
    m_l^Z = \sum_{\bs_j\in\bs_l^Z}m_j
\end{equation}

Zooming has three effects on the risk measure. Firstly, the center of a minimal $k$-anonymity region can not be closer to the unit than the zoom-factor allows. This can be taken into account by making the $\delta$ parameter depending on the zoom-factor. For example, we could choose $\delta(Z)$ equal to the absolute size of the pixels at the used zoom-level or we could choose $\delta(Z) = \sup \{\dist{\bs_i-\bs_j}: \pi_Z(\bs_i)=\pi_Z(\bs_j)\}$, being the largest $L_2$ distance between visually overlapping locations in $\S$.

Secondly, the count function then has to take into account the multiplicity of the zoomed-locations. That is, we need to use 
\begin{equation}
    \NSZ(B(\bx,r)) = \sum_{l=1}^{L_Z} m_l^Z \mathbb{1}([\bx]_Z\in\pi_Z(B(\bx,r)))
\end{equation}

Thirdly, the radius needs to be transformed to the zoom-resolution. I.e., the radius no longer is defined in meters but in pixels.
Essentially, the risk measure then measures the \emph{perceived} risk when publishing at zoom-level $Z$: zooming reduces the perceived (in the publication visible) risk by overlapping observations, but the underlying intrinsic risk does not change.

For the sake of readability, we will henceforth focus on the measures as defined in (\ref{eq:Fkdelta}) and (\ref{eq:Rkdelta}). The discussions can however be extended to the versions where the zoom effect is taken into account.

\subsection{Special cases}
\subsubsection{Point-oriented}
Taking $\delta=0$ results in looking for the smallest $k$-anonymity region that contains the unit of interest and has its center at the location of that unit. This is what we call a point-oriented risk measure and is closely related to the notion of $k$-nearest-neighbors. It also intuitively resembles the standard interpretation of regional $k$-anonymity: it measures the local density of the population around the unit of interest.

One drawback of this measure becomes visible when the target unit is close to the border of a populated area. Since the $k$-anonymity area is symmetric around the unit, all units that contribute to the total number of units in the $k$-anonymity region have to come from within the populated area. Hence the minimal radius can be quite large. This can be seen as an artifact of this version of the measure, that appears near the borders.
\subsubsection{Cluster-oriented}
Taking $\delta=\infty$ results in looking for the smallest $k$-anonymity region that contains the point of interest and at least $k$ units. This is related to finding an anchored, weighted minimal enclosing circle of a set of points. Anchored in the sense that the unit of interest needs to be inside the circle, weighted in the sense of the restriction on the number of \emph{units} instead of locations and minimal enclosing circle in the sense of the smallest radius. The cluster-oriented approach is also related to $k$-center clustering. 

With this measure units near the border of populated areas can get smaller $k$-anonymity areas: shifting the center of the circle into the populated area usually yields a smaller radius compared to the point-oriented version.

\section{Example}\label{sec:example}
To illustrate the risk measure defined in (\ref{eq:Rkdelta}), we will use the enterprises dataset that is included in the R-package \texttt{sdcSpatial}~\cite{JongeWolf2025}. This dataset contains fake, but realistic locations of enterprises and some simulated data per location. The total number of enterprises is 8348, the number of unique locations is 8055, so this dataset is indeed a multiset. The distribution of the multiplicities is given in Table~\ref{tab:multiset}.
\begin{table}[htb]
\newdimen\mywidth
\setbox0=\hbox{7945}
\mywidth=\wd0
\centering
\caption{Number of locations per multiplicity in enterprises dataset}\label{tab:multiset}
    \begin{tabular}{l*{11}{R{\mywidth}}}
        Multiplicity & 1 & 2 & 3 & 4 & 5 & 8 & 9 & 10 & 23 & 25 & 68 \\
        Locations & 7945 & 80 & 9 & 10 & 3 & 1 & 3 & 1 & 1 & 1 & 1
    \end{tabular}
\end{table}

\subsection{Point-oriented}
Finding the minimal $k$-anonymity region for $\delta=0$ for a given target location $\bs$ is relatively easy. Since $\bs$ needs to be the center of the circle, the smallest circle contains those locations of the $k$ nearest neighbors of $\bs$ such that the total multiplicity is at least $k$. Pseudo code for this approach is given in Algorithm~\ref{alg:one}.

\begin{algorithm}[h!tb]\label{alg:one}
\SetAlgoLined
    \emph{numunits} $=$ multiplicity of $\bs$\\
    \emph{bestradius} $=0$\\
    \If{numunits < $k$}{
        find $k$ nearest neighbor locations of $\bs$ with at least one unit\\
        \While{numunits < $k$}{
            add multiplicity of next nearest neighbor location to \emph{numunits}
        }
        \emph{bestradius} $=$ distance between $\bs$ and last added neighbor location
    }
    \KwRet{bestradius}
\caption{Radius of minimal $k$-anonymity region for $\delta=0$}
\end{algorithm}

\begin{figure}[h!tbp]
\centering
\includegraphics[width=\textwidth]{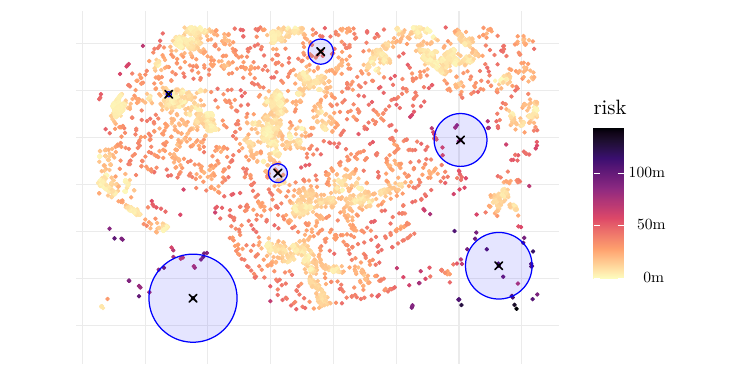}
\caption{Point-oriented risk; $\times$ denotes a selected location with corresponding minimal circle}\label{fig:point}
\end{figure}

Applying this to the enterprises dataset and plotting the scaled risk of equation~(\ref{eq:Rkdelta}) with $k=10$, we get Figure~\ref{fig:point}. For six locations we added the corresponding minimal circle for illustration.

\subsection{Cluster-oriented}
Finding the minimal $k$-anonymity region for $\delta=\infty$ is a combinatorial optimization problem and sometimes referred to as a weighted, anchored minimal enclosing circle problem. At first one might be tempted to think that, in case of all multiplicities equal to 1, such minimal enclosing circle will be the smallest circle surrounding the $k-1$ nearest neighbors of location $\bs$. However, this is not necessarily the case as the following counterexample shows. 

\begin{figure}[h!tb]
\centering
\begin{tikzpicture}[scale=0.75]
    \draw[help lines, color=gray!30, dashed] (-3,-3) grid (3,3);
    \fill (0,0) circle (3pt) node[above left] {$\bs_1$}
          (2,0) circle (3pt) node[above right] {$\bs_2$}
          (-2,0) circle (3pt) node[below left] {$\bs_3$}
          (0,-2.5) circle (3pt) node[below left] {$\bs_4$};
    \draw[blue,dashed] (0,0) circle (2);
    \draw[red] (1,-1.25) circle (1.601);
    \draw (-2.9,0)--(2.9,0);
    \draw (0,-2.9)--(0,2.9);
\end{tikzpicture}
\caption{\emph{Counterexample}}\label{fig:example3}
\end{figure}
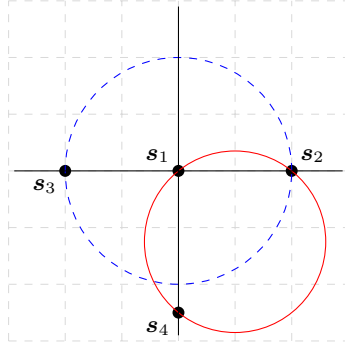

\noindent\hrulefill\\
\noindent\emph{As a counterexample, examine Figure~\ref{fig:example3}, with the four points $\bs_1=(0,0)$, $\bs_2=(2,0)$, $\bs_3=(-2,0)$ and $\bs_4=(0,-2.5)$. 
The nearest neighbors of $\bs_1$ are obviously $\bs_2$ and $\bs_3$. 
The radius of the minimal enclosing circle containing these three point equals 2 (the blue, dashed circle in Figure~\ref{fig:example3}).
However, the radius of the minimal enclosing circle containing points $\bs_1$, $\bs_2$ and $\bs_4$ equals $\sqrt{10.25}/2 \approx 1.6$. 
Hence, the smallest circle containing $\bs_1$ and at least two other points is defined by $\{\bs_1,\bs_2,\bs_4\}$ (the red, solid line circle in Figure~\ref{fig:example3}).}

\noindent\hrulefill\\
\noindent In theory, the complete set of locations needs to be analyzed. However, one may start with examining locations in the neighborhood of the target location and analyze all possible circles with at least $k$ units using the locations in that neighborhood. Finding the weighted, anchored minimum enclosing circle among those neighboring locations gives an upperbound $r_0$ for the minimum over all locations. Using other points can only lower that upperbound if such a point is not further away than $2r_0$. This reduces the set of points that needs to be considered. Particularly in case $k \ll \NS(\mathbb{R}^2)$ this reduces the calculation time considerably. Pseudo code for this approach is given in Algorithm~\ref{alg:two}.

\begin{algorithm}[h!tb]\label{alg:two}
\SetAlgoLined
    \emph{numunits} $=$ multiplicity of $\bs$\\
    \emph{bestradius} $=0$\\
    \If{numunits < $k$}{
        \mcc{Step 1: find valid initial upperbound}
        $S =$ set of nearest neighbors of $\bs$ with total muliplicity of at least $k$\\
        \emph{bestradius} $=$ radius of minimal enclosing circle of $S \cup \{\bs\}$\\
        \mcc{Step 2: find optimal radius}
        \emph{candidates} $=$ all points within distance $2\times$\emph{bestradius} from $\bs$\\
        \For{each support set $D$ from candidates with size 1, 2, or 3}{
            Calculate circle $C$ defined by support set $D$\\
            \If{$($multiplicity of $C>=k)$ \emph{\&} $(\bs\in C)$ \emph{\&} $($radius$(C)$ $<$ bestradius$)$}
                {\emph{bestradius} $=$ radius$(C)$}
        }
    }   
    \KwRet{bestradius}
\caption{Radius of minimal $k$-anonymity region for $\delta=\infty$}    
\end{algorithm}

\begin{figure}[h!tb]
\centering
\includegraphics[width=\textwidth]{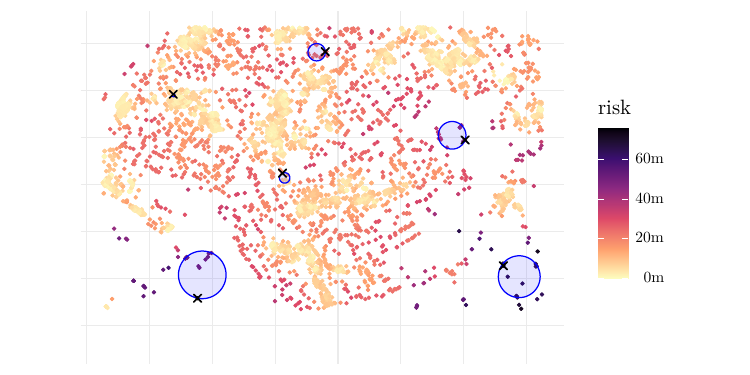}
\caption{Cluster-oriented risk; $\times$ denotes a selected location with corresponding minimal circle}\label{fig:cluster}
\end{figure}

Applying this to the enterprises dataset and plotting the scaled risk of equation~(\ref{eq:Rkdelta}) with $k=10$, we get Figure~\ref{fig:cluster}. For the same six locations as in Figure~\ref{fig:point} we added the corresponding minimal circle for illustration.

\section{Discussion}\label{sec:discussion}
We defined a risk measure as the minimal circular $k$-anonymity region for a target unit, with the idea that the unit at the target location is then in some sense indistinguishable from the other units within that $k$-anonymity region. An alternative interpretation would be that the units in that $k$-anonymity region influence the value of some attribute of the target unit when using a (kernel) density map.

In thematic maps published by statistical offices, the set of locations of the units of the (target) population is often a multiset. The definition of our risk takes that into account by counting the number of units instead of the number of locations. As a consequence, the minimal $k$-anonymity region may encapsulate less than $k$ unique locations, while still containing at least $k$ units. Indeed, the region may even contain more than $k$ units. The scaling of the measure (as in equation~\ref{eq:Rkdelta}) links the risk measure to the local density of the population.

Since our risk measure is linked with the target location and does not depend on the regional classification used in the publication, it does not suffer from the Modifiable Areal Unit Problem. Moreover, the fact that we are using circles implies that the $k$-anonymity regions are rotation invariant. It could be interesting to investigate how this risk measure would look like and how it would behave when using a `ball' under a different metric (e.g., under $L_1$ metric a ball would be a square).

The resolution of a published thematic map induces a perceived risk: at a low resolution some in real life unique locations may seem to overlap. We showed how to adjust our risk measure to take into account the perceived disclosure risk. When zooming in or out while displayin a thematic map, the screen resolution effectively changes. Adjusting our risk measure as described in subsection~\ref{subsec:zooming}, could take that into account and could even suggest a maximum (most detailed) zoom level such that the perceived disclosure risk does not exceed a certain threshold. 

The $k$-anonymity region could also be used as a `natural' region to use in calculating how much a unit contributes to the overall value of a region. For example, one could calculate a $p\%$-rule on the values of the units in the $k$-anonymity region. That way it is possible to incorporate attribute disclosure more explicitly.

The example in section~\ref{sec:example} shows some of the properties of our risk measure we discussed in the section where we defined the measure. In Figure~\ref{fig:point} we see at the lower left corner that locations near the edge of a populated region get a seemingly large minimal $k$-anonymity region. This is the consequence of fixing the target location at the center of the $k$-anonymity region. Letting go of the restriction of centering the $k$-anonymity region at the target location, reduces the radius of the minimal $k$-anonymity region, as can be seen at the lower left corner in Figure~\ref{fig:cluster}. This shows that the cluster-oriented version is less affected by artifacts near the boundary of populated areas compared to the point-oriented version. The same examples show the monotonicity of the risk measure as function of $\delta$: the discs for the cluster-oriented version ($\delta=\infty$) are smaller or equal to the ones for the point-oriented version ($\delta=0$).

One direction for further research should aim at connecting our risk measure to kernel density maps. As mentioned, the risk measure is related to the amount units' values may influence the target unit when using a kernel density map. It is interesting to research the relationship of our risk measure with the `risk-optimal' bandwidth of a kernel density estimator.

Another direction could be to investigate how this risk measure could be used in publishing tabulated data instead of thematic maps. In tabulated data region can play a role as well. However, in that case it might be interesting to study the risk measure with an additional restriction that the $k$-anonymity region should be enclosed within or restricted to the regions of the tabulated data.

Yet another direction could be to investigate how this risk measure behaves when applying methods to control the disclosure risk. Moreover, we should research whether (and how) this measure can be used to adjust disclosure control measures such that they take into account the spatial characteristics of the units.

\bibliographystyle{splncs04}
\bibliography{psd2026}

\end{document}